Dispersion Relations for Thermally Excited Waves in Plasma Crystals

K. Qiao and T. W. Hyde
Center for Astrophysics, Space Physics and Engineering Research
Baylor University
P.O.Box 97310, Waco, TX. 76798-7310, USA

Abstract

Thermally excited waves in a Plasma crystal were numerically simulated using a Box_Tree code. The code is a Barnes_Hut tree code proven effective in modeling systems composed of large numbers of particles. Interaction between individual particles was assumed to conform to a Yukawa potential. Particle charge, mass, density, Debye length and output data intervals are all adjustable parameters in the code. Employing a Fourier transform on the output data, dispersion relations for both longitudinal and transverse wave modes were determined. These were compared with the dispersion relations obtained from experiment as well as a theory based on a harmonic approximation to the potential. They were found to agree over a range of $0.9<\kappa<5$, where $\kappa$ is the shielding parameter, defined by the ratio between interparticle distance $a$ and dust Debye length $\lambda_D$. This is an improvement over experimental data as current experiments can only verify the theory up to $\kappa = 1.5$.

1. Introduction

There has been increasing interest in the field of Complex plasmas for the past two decades, especially after Coulomb crystals were first experimentally observed at various labs [1-3]. Recently, the study of waves in Plasma crystals has become one of the more active research fields, since as in condensed matter systems, wave propagation is a research topic with both theoretical and practical value.

Experimental research has been conducted [4-8] on both longitudinal and transverse waves in 2D Coulomb crystals. In these experiments, a 2D hexagonal lattice of dust particles was established within the sheath region of the lower electrode in a rf discharge plasma. Dust particles introduced into the plasma become charged and are subject to mutual Coulomb interactions. The particles interact with each other through a Yukawa potential, $V(r) = (Q/r)\exp(-r/\lambda_D)$, where $\lambda_D$ is the Debye length, which helps determine the manner in which the charge will be shielded by the plasma. The particles are levitated by an electric field in the sheath region and make a 2D Coulomb crystal. The Coulomb crystal is primarily characterized by two parameters, the shielding parameter $\kappa=a/\lambda_D$ and the coulomb coupling parameter $\Gamma=Q^2/4\pi\varepsilon_0 a k_B T$, where $a$ is the mean interparticle spacing and T is the kinetic temperature of the particles.

Several laboratory experiments have determined that exciting particles in a shaped region [4-7] of the lattice can produce both longitudinal and transverse waves. In these experiments, particle excitation was created using a sinusoidally modulated laser beam. Under the radiation pressure created by the laser, the particles within the excitation region oscillate sinusoidally, and a wave propagates through the lattice. Fourier analysis of the subsequent particle motion allows the dispersion relations to be determined. More recently, experiments [8] examining waves created via thermal excitation have been conducted. The dispersion relations for both the longitudinal and transverse waves were again determined employing Fourier analysis of the spontaneous motion of the particles within the lattice. Comparison of the results of the two experiments show that they agree with one another, even though the results from thermally produced waves cover a larger region in k space.

A theoretical model has recently been given [9] to explain these experimental results. X. Wang et al. analytically derived the dispersion relations for both the longitudinal and transverse waves on a 2D hexagonal lattice assuming a harmonic approximation for the motion of the particles about their equilibrium position and a Yukawa interparticle potential. Wang et al. showed that not only do the dispersion relations depend upon the direction of propagation but that waves propagating parallel to the primitive translation vectors have different dispersion relations than those propagating perpendicular to the primitive translation vectors. In both cases, the dispersion relations were shown to agree with experiment.

## 2. Methods

In this research, a Box_Tree code was employed to simulate the spontaneous thermal wave in a Coulomb crystal. The results were Fourier analyzed to obtain the dispersion relations and then compared with both the theoretical results discussed above and experimental data.

The Box_Tree code is a Barnes-Hut tree code first written by Derek Richardson [10] for planetary ring and disk studies. It was later modified by Lorin Matthews [11] to include electrostatic interactions and then by John Vasut [12] to simulate the formation of plasma crystals. Box_Tree is proving to be an effective tool for modeling real time systems, particularly those composed of large numbers of particles with specific interparticle interactions.

The Box_Tree code models a dusty plasma system by first dividing it into self-similar patches, where the box size is much greater than the radial mean excursions of the constituent dust grains and the boundary conditions are met using twenty-six ghost boxes. A tree code is incorporated into the routine to allow it to deal with interactions between the particles. Interparticle interactions are calculated by examining the multipole expansions of collections of particles, so the code scales as $N \cdot \log N$ instead of $N^2$. Data files showing each particle's position and velocity are output for analysis once a certain user specified time interval is reached. Particle charge, mass, density, average initial velocity, Debye length and output data intervals are all adjustable parameters within the code.

All interparticle interactions include both gravitational and electrostatic forces. The interparticle electrostatic force is assumed to be a 3D screened Coulomb repulsion or Yukawa potential. The simulation box is a 3D box and all calculations are handled in a

3D manner. However, the initial positions of the particles are constrained to lie in a horizontal plane (as is often the case in an experimental setting) and their initial velocities are specified to contain only XY components. The gravitational force from the earth and the electrostatic force produced by the (lower) powered electrode are not assumed to balance one another and are neglected. Therefore, the particles are restricted to a 2D plane (within the 3D box) throughout the simulation since they are not acted upon by forces (or velocities) in the Z direction.

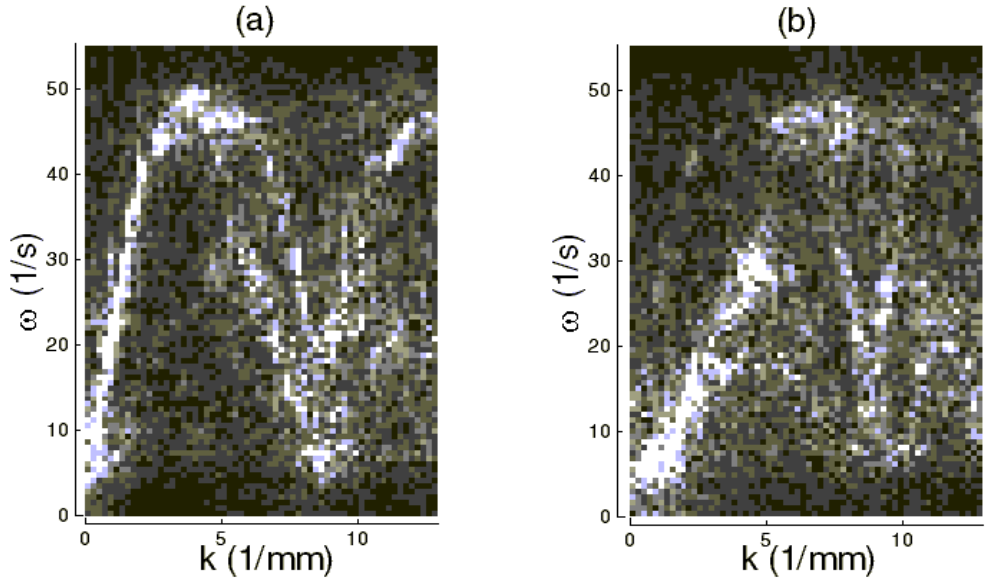

Fig. 1: Dispersion relations from the simulation of (a) longitudinal and (b) transverse waves with an arbitrary propagation direction in a Coulomb crystal with $\kappa = 1.4$ ($\lambda_D = 570$ μm).

Initial conditions include a random distribution of 679 particles in a 20×20×20mm$^3$ box. The neutral gas drag is included with a specified Epstein drag coefficient of $\gamma =1.22$. After about 20 seconds, the state of the system was determined to be in crystalline form (solid) using the pair correlation function. The pair correlation function was employed (along with ordering and Voronoi diagrams) instead of the usual phase space diagrams often seen in MD simulations since the kinetic energy (temperature) of the particles will continue to decrease throughout the simulation due to gas drag. This manifests itself in a cooling of the system making the usual phase space diagrams difficult to use due to their time dependence (Ref. 12). In all of the simulations, the diameter of the embedded particles is 6.5 μm, the density is 1.51 g/cm$^3$, the charge is 14,500 e and the interparticle distance as determined from the pair correlation function is around 825 μm.

The number of particles (or the box size) will limit the resolution of any subsequent graphs but the minimum wavelength is only determined by the width of the bins chosen to analyze the data. For best results, the binwidth must be comparable to the minimum interparticle spacing along the specified propagation direction. Waves with wavelengths shorter than the minimum interparticle spacing are not detectable.

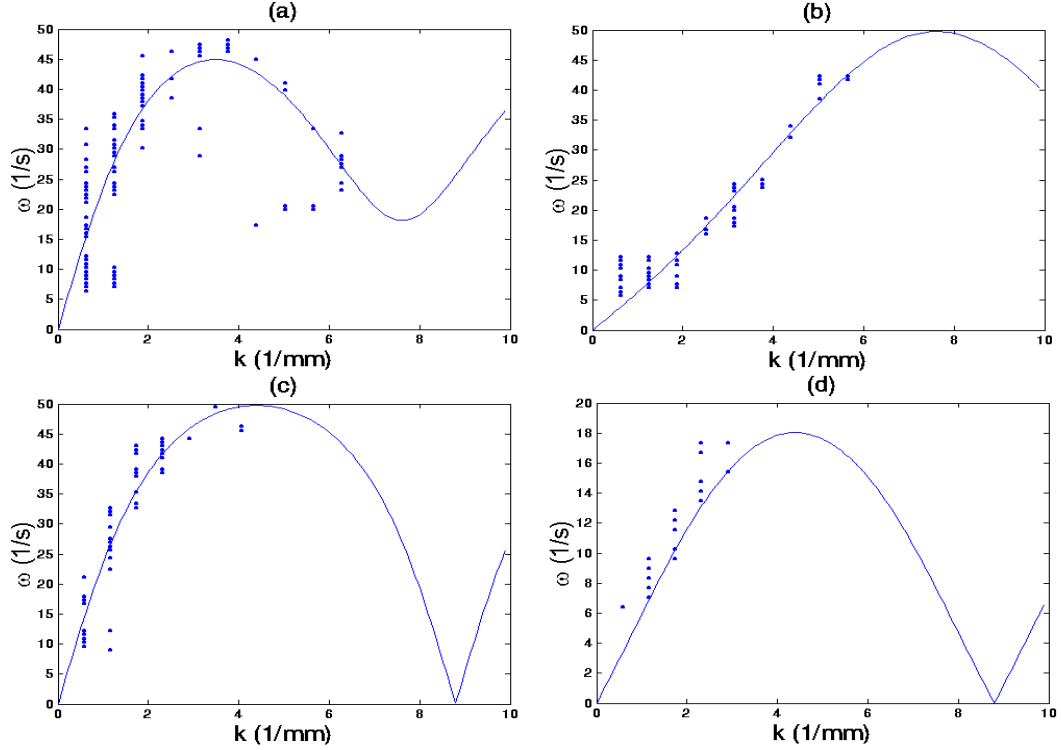

Fig. 2: Comparison of dispersion relations as obtained from simulation and Wang's theory. Dots represent data obtained from simulation. Solid lines are dispersion relations from Wang's theory. In all cases, κ = 1.4 ($\lambda_D$ = 570 μm).

The possibility of temperature fluctuations of the system (temperature control / system thermostat) is carefully monitored and has been shown to be negligible for all conditions examined by this simulation. (See [12] for complete details of this method.) Briefly, the initial velocities are specified with a user defined average value and a Gaussian distribution. From them, the system temperature is then calculated and any temperature fluctuations compared to the dust temperature to assure they are negligible. The dispersion relations were also examined for various temperatures to ensure that they did not depend on the overall system temperature.

Once the coulomb lattice is established, the random thermal motion of the particles within this lattice is tracked for 10 seconds by choosing the data output time interval to be 0.01 second and collecting 1000 data files. Depending on the equilibrium particle position, the particles are divided into bins with the particle velocity averaged over each bin, yielding velocity data, which is dependent upon position. Combining data files, a velocity matrix depending on time and position can then be obtained using a double Fourier transformation of this matrix as given below. This yields a matrix representing the particle velocity in space.

$$V_{k,\omega} = 2/TL \int_0^T \int_0^L v(x,t)\exp[-i(kx-\omega t)]dxdt$$

## 3. Results

Fig. 1 shows this matrix with pixel brightness corresponding to velocity value and the bright areas showing the dispersion relation of the (a) longitudinal and (b) transverse spontaneous waves for $\kappa = 1.4$ ($\lambda_D = 570$ μm) and particle charge q $=1.92*10^{-15}$ C. To obtain the dispersion relation for a particular wave mode (longitudinal or transverse), velocity components perpendicular to the bins for longitudinal waves and parallel to the bins for transverse waves must be used.

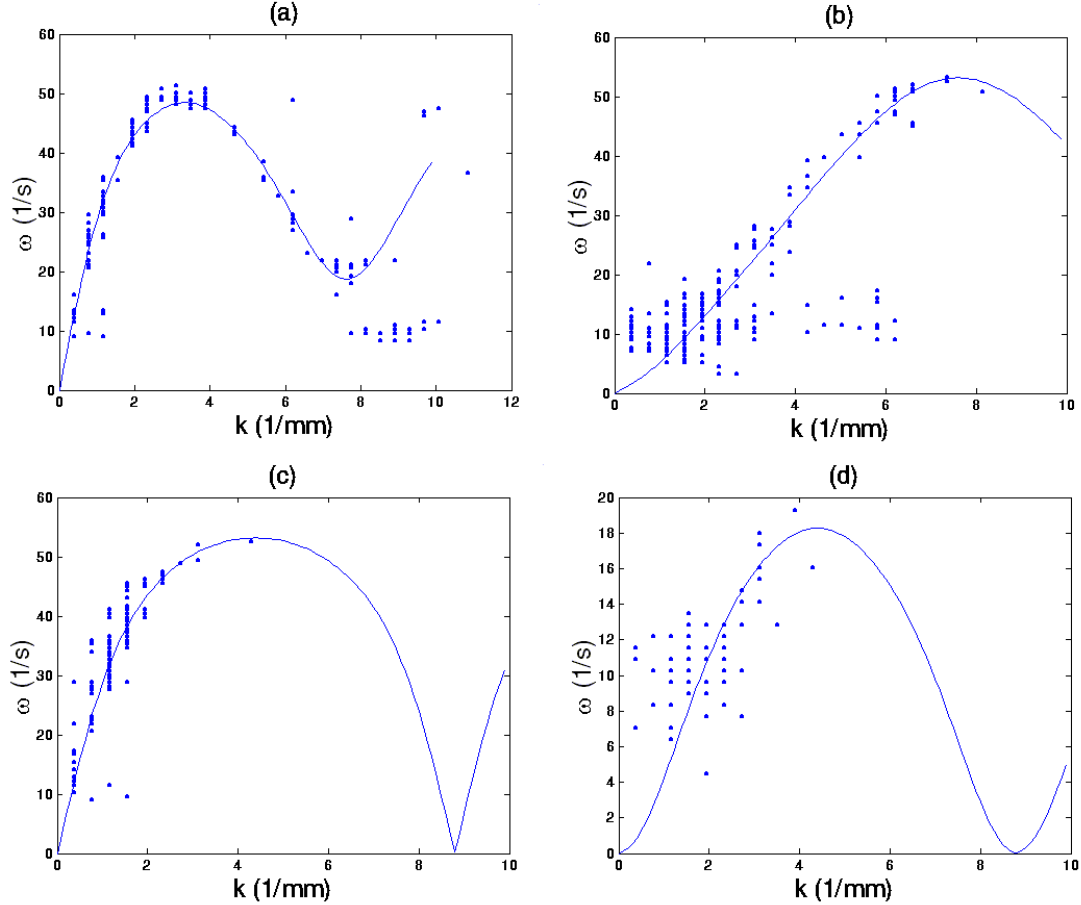

Fig. 3: Same as for Figure 2 but with $\kappa = 0.9$ ($\lambda_D =0.92$mm).

Fig. 1 was obtained for arbitrary propagation direction. To choose a particular propagation direction, the bins should be chosen perpendicular to that specified direction.

The shape of the dispersion relation shown in Fig. 2 is obtained by setting a minimum intensity level on the data in a figure similar to Fig.1 but with specified propagation directions. Superimposing the dispersion relation obtained from Wang's theory as a solid line, it can be seen that the raw simulation data is a solid fit. As mentioned above, the possible wavelengths observed are limited by the width of the bins, which is in turn limited by the interparticle spacing. Thus, the maximum wave number k is about 4 mm$^{-1}$ for propagation perpendicular to the primitive translation vector and 6 mm$^{-1}$ for propagation parallel to the primitive translation vector. The finite width of the bins is also responsible for longitudinal dispersion relation data points sometimes falling along the transverse dispersion relation, and vice versa. As can also be seen, for example, in Fig. 3 (b), sometimes the data points from simulations of waves propagating perpendicular to the primitive translation vector agree with the theoretical dispersion relation of waves propagating parallel to the primitive vector, and vice versa. This is because the Coulomb crystal examined is not an ideal solid lattice.

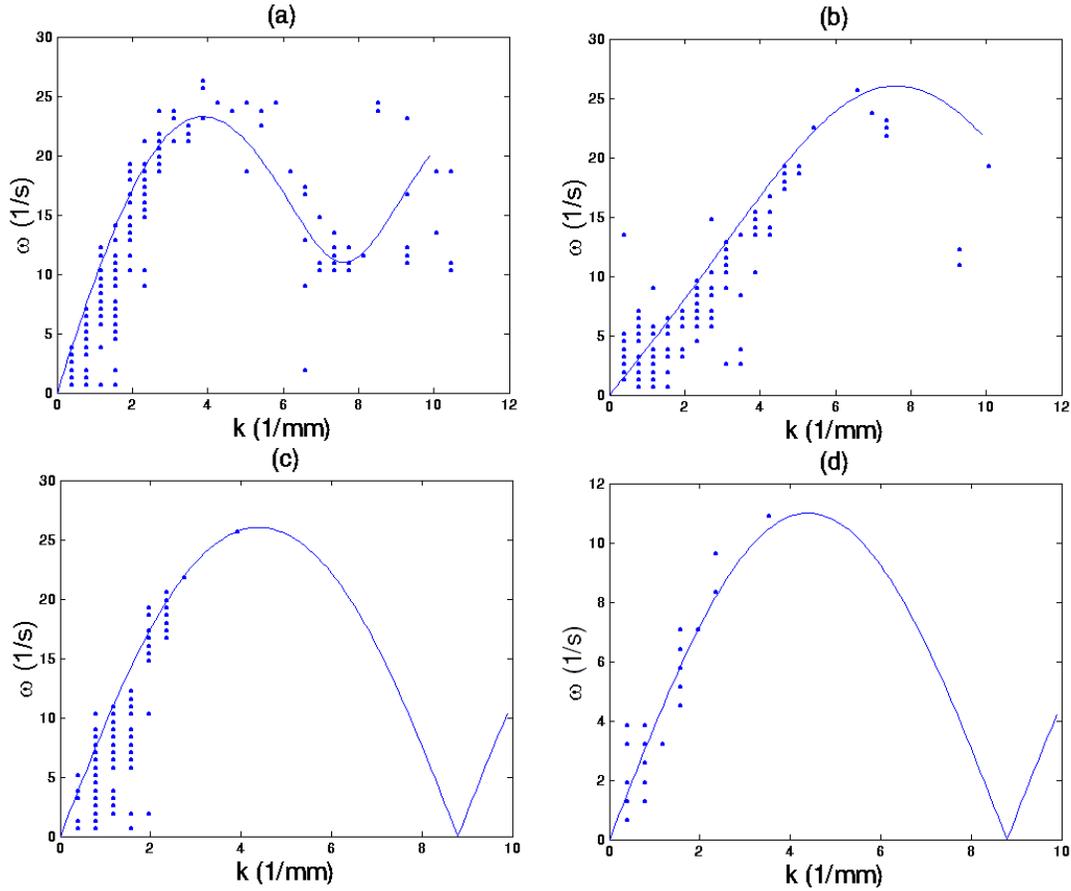

Fig. 4: Same as for Figure 2 but with $\kappa = 5$ ($\lambda_D = 0.17$mm).

The simulation was run for different average initial velocities ranging from 0 to $1.0\times10^{-3}$ m/s. The dispersion relations were seen to be almost identical, showing them to be temperature independent.

The simulation was run also for different $\kappa$ values, over the regime $0.9<\kappa<5$. The results for $\kappa = 0.9$ ($\lambda_D = 0.92$mm) and $\kappa = 5$ ($\lambda_D = 0.17$mm) are shown in Figures 3 and 4. The dispersion relations obtained from the raw data were again found to be in agreement with the theory for both longitudinal and transverse waves. For values of $\kappa > 5$, the dispersion curve is too low and the shape is no longer clear enough to obtain distinguishable data.

## 4. Conclusions

As can be seen, Box_Tree produces dispersion relations, which are in excellent agreement with current theory and can be run over a larger range of $\kappa$ than is experimentally possible. This allows Box_Tree to be used as an effective numerical tool for simulation of Yukawa systems. System parameters are much easier to adjust than in an experimental setting and can cover regimes (such as for high frequencies) which are difficult (if not impossible) to reach physically.

Finally, the code is not predicated on the assumptions of a harmonic approximation and/or lattice structure. The only assumptions required by the program are a Yukawa potential and the specification of constant parameters (i.e., charge, mass, Debye length, etc) by the user. This should allow Box_Tree to become a effective tool in simulating nonlinear effects or complex plasma liquids.